\documentclass[aps,prd,twocolumn,superscriptaddress,noshowpacs,showkeys,
nofootinbib
]{revtex4}
\usepackage{graphicx,epsfig}
\usepackage{amsmath,amssymb}
\usepackage{lineno,hyperref}
\hypersetup{colorlinks=true}
\usepackage{graphicx}
\usepackage{wrapfig}
\usepackage{subfigure}
\modulolinenumbers[5]

\renewcommand{\vec}[1]{\ensuremath{\mathbf{#1}}} 
 
\newcommand{\dd}{\mathrm{d}} 
\newcommand{\ii}{\mathrm{i}} 
 
\newcommand{\eee}[1]{\mathrm{e}^{#1}}

\newcommand{\pd}[2]{\frac{\partial #1}{\partial #2}} 
 
\renewcommand{\phi}{\varphi}

\hypersetup{
 colorlinks=true,
 linktoc=all,
 linkcolor=red,
 citecolor=blue,
 urlcolor = blue}

\begin{document}

\title{Emergence of Fresnel diffraction zones in gravitational lensing by a
cosmic string}

\author{Isabel Fern\'{a}ndez-N\'{u}\~{n}ez}
\affiliation{Departament de F\'{i}sica Qu\`{a}ntica i Astrof\'{i}sica}
\affiliation{Institut de Ci\`{e}ncies del Cosmos (ICCUB) \\ Facultat de
F\'{i}sica, Universitat de Barcelona, Mart\'{i} i Franqu\`{e}s 1, E-08028
Barcelona, Spain.}
\author{Oleg Bulashenko}
\affiliation{Departament de F\'{i}sica Qu\`{a}ntica i Astrof\'{i}sica}

\date{\today}

\begin{abstract}
The possibility to detect cosmic strings -- topological defects of early
Universe, by means of wave effects in gravitational lensing is discussed.
To find the optimal observation conditions, we define the hyperbolic-shaped
Fresnel observation zones associated with the diffraction maxima and analyse
the frequency patterns of wave amplification corresponding to different
alignments.
In particular, we show that diffraction of gravitational waves by the string
may lead to significant amplification at cosmological distances.
The wave properties we found are quite different from what one would expect,
for instance, from light scattered off a thin wire or slit, since a cosmic string,
as a topological defect, gives no shadow at all.
\end{abstract}



\maketitle

\section{Introduction}

The first direct detection of gravitational waves by the Laser Interferometer
Gravitational-Wave Observatory (LIGO) \cite{LIGO16-1} opened up a new way to
observe the Universe. Along with gravitational wave detection, it was the first
direct observation of binary black holes.
With this success, there are many hopes that other previously invisible
cosmological objects, which emit or scatter gravitational waves, will be
observed in the near future.

In this paper we discuss the possibility to detect cosmic strings --
topological defects that may have been formed in the early Universe
\cite{kibble76, vilenkin-shellard94} -- by means of wave effects in the
gravitational lensing taking into account the interference and diffraction.
We emphasize the difference of wave diffraction on a topological defect from
that on a compact object.
For the wave effects to be detectable in a compact-mass gravitational lens, the
wavelength $\lambda$ should be comparable or larger than the Schwarzschild
radius $R_{\rm s}$ of the lens \cite{deguchi86}.
In this case, the Fresnel number, which is the key parameter for the
diffraction, is given by $R_{\rm s}/\lambda$, and the diffraction scales like
$O(\lambda/R_{\rm s})$.
This scaling cannot be applied to a string, a non-compact object with conical
topology.
It has been shown recently for the plane-wave diffraction by string
\cite{pla-string16} that the Fresnel number is determined by the ratio
$r \Delta^2/\lambda$, where $r$ is the distance from the string to the observer
and $\Delta$ is a constant related to the deficit angle of conical space, which
is proportional to the linear mass of the string
\cite{vilenkin81,vilenkin84,gott85}.
For the typical $\Delta\sim 10^{-7}$,
low Fresnel numbers can be achieved at cosmological distances from the
string, $r\sim 10^{14\,}\lambda$.
As a result, the diffraction effects can be of the same order as the geometrical
optics giving an additional amplification at the observation point
\cite{pla-string16}.
This is a direct consequence of the conical topology, for which the metric is
locally flat, but globally it forces the parallel geodesics to cross (when they
pass on opposite sides of the string) at a large distance.
On the other hand,  the deflection angle, equal to $\Delta$, is independent of
the impact parameter \cite{vilenkin81,vilenkin84}. Hence, the characteristic
fringe width in the interference pattern $\sim \lambda/(2\Delta)$ does not vary
with distance.
This is another feature distinct from the compact-object lens, for which the
interference fringe scales with distance as $\sim \lambda\sqrt{r/R_s}$
\cite{nakamura98}.

The objective of this paper is twofold. First, we study the question of how the
Fresnel diffraction zones emerge under wave propagation in conical spacetime
created by a straight cosmic string \footnote{Actual strings are not straight and may contain loops, we refer to a
straight-line segment of an infinitely long or closed string lying at the
observer-source line of sight.}.
The diffraction pattern we have obtained is quite different from what one would
expect from light scattered off a thin wire or slit
\cite{sommerfeld54,born-wolf-03}, since the cosmic string, as a topological defect,
gives no shadow.
After an appropriately chosen coordinate transformation, we convert the problem
of a single-source wave in conical space to a more tractable form with a locally
Minkowskian line element and a limitation on the angular range.
As a result, we obtain the interference and diffraction pattern analytically as
a superposition of wave fields from two image sources illuminating two virtual
half-plane screens.

Second, we take into account the curvature of the incident wavefront by
considering the wave source at a finite distance from the string. This is a
more general case with respect to our previous study \cite{pla-string16}.
By applying the uniform asymptotic theory of diffraction \cite{boersma68,borovikov},
we obtain analytical solutions for the wave field in the whole space including
the lines of singularities at the boundaries of the double-imaging region.
Away from the boundaries, the wave field is interpreted in the framework of
Keller's geometrical theory of diffraction \cite{keller62},
which has demonstrated to be quite efficient in studying diffraction on a
topological defect \cite{pla-string16}.
Our results allow to predict with high accuracy the location of the diffraction
maxima both in coordinate space and in energy spectrum, along with the nodal
and antinodal lines of geometrical-optics interference. 
We found it convenient to associate the diffraction maxima with what we call the
``Fresnel observation zones'', that help to localize the regions where the
amplification due to the string is the highest and easier to observe.
The boundaries between the zones are determined by hyperbolas in an equivalent
Minkowskian space.
In the limit of an infinitely distant source (incident plane wave), the hyperbolas
convert to parabolas, all with a common focus at the string.

\section{Wave equation in conical spacetime} \label{sect:wave-eq}

We start with a spacetime metric for a static cylindrically symmetric cosmic
string \cite{vilenkin81,gott85}
\begin{equation}
\dd s^2= -\dd t^2 + \dd r^2 + (1-4G\mu)^2r^2\dd \phi^2 + \dd z^2,
\label{eq:metric}
\end{equation}
where $G$ is the gravitational constant, $\mu$ is the linear mass density of the
string lying along the $z$-axis, ($t,r,\phi, z$) are cylindrical coordinates,
and the system of units in which the speed of light $c=1$ is assumed.
With a new angular coordinate $\theta=(1- 4G\mu) \phi$,
the metric \eqref{eq:metric} takes a locally Minkowskian form
\begin{equation}
\dd s^2= -\dd t^2 + \dd r^2 + r^2\dd \theta^2 + \dd z^2,
\label{eq:metric-Mink}
\end{equation}
having, however, a limitation on the angular range. It is assumed here, that 
a wedge of angular size $8\pi G\mu$ is taken out and the two faces of 
the wedge are identified \cite{vilenkin81,vilenkin-shellard94}.
By introducing the deficit angle $2\Delta$ with
\begin{equation}
\Delta = 4\pi G \mu,
\label{eq:delta}
\end{equation}
the angular coordinate $\theta$ spans the range $2\pi - 2\Delta$.

\begin{figure}[t]
\centering
\includegraphics[width=0.75\columnwidth]{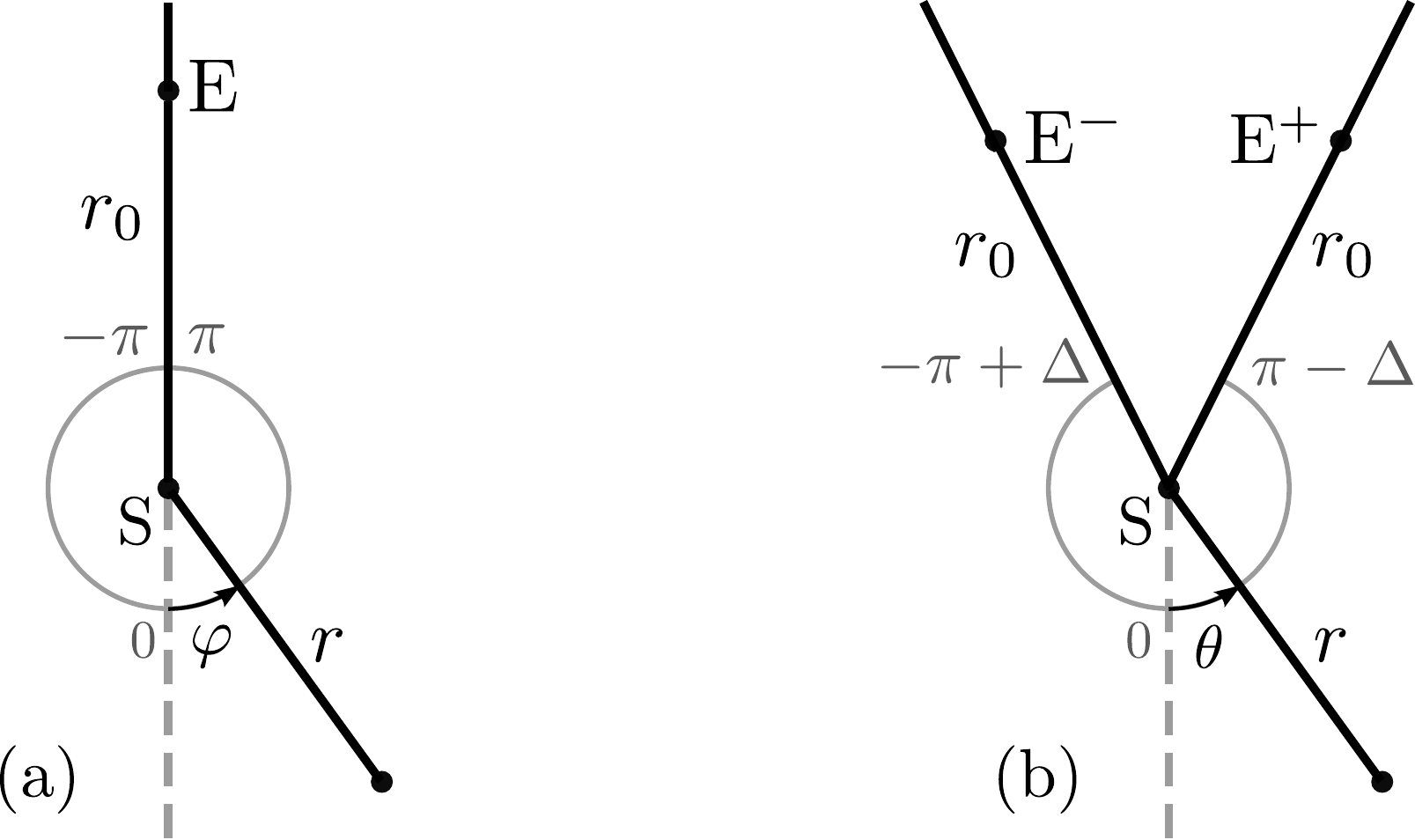}
\caption{
Geometry of conical space at the $z=0$ plane for two equivalent backgrounds with
point $S$ indicating the location of the string:
(a) polar coordinates $(r,\phi)$ with a source $E$;
(b) Minkowskian coordinates $(r,\theta)$ with deficit angle $2\Delta$ and two image sources
$E^-$, $E^+$.}
\label{fig:split}
\end{figure}

We consider the question of finding a solution of the wave equation in
background \eqref{eq:metric} corresponding to a time harmonic source,
situated at a finite distance from the string.
For the sake of simplicity, in order to keep the problem two-dimensional, we
consider a line source parallel to the string.
Our aim is to see how a wave emitted by a line source is diffracted in conical
spacetime. The wave equation in background \eqref{eq:metric} for the scalar
field $U(r,\phi)$ is (see, e.g., \cite{pla-string16,linet86,suyama06})
\begin{equation}
\left(
\pd{^2}{r^2}+\frac{1}{r}\pd{}{r}+\frac{1}{\beta^2 r^2}\pd{^2}{\phi^2}
+\omega^2 \right)U=0,
\label{wave-eqn}
\end{equation}
where we denoted $\beta\equiv 1-\Delta/\pi$.
We assume that Eq.\ \eqref{wave-eqn} is valid for electromagnetic waves,
as well as for gravitational waves (in an appropriately chosen gauge) when the
effect of gravitational lensing on polarization is negligible and both types of
waves can be described by a scalar field \cite{misner}.
Consider a line source $E$ located at $\vec{r_0}=(r_0,\pi)$ and emitting a
cylindrical wave described by
\begin{equation}
U=A \,H_0^{(1)} (k |\vec{r}-\vec{r_0}|),
\end{equation}
where $A$ is a normalization constant and $H_0^{(1)}$ is the Hankel function of
the first kind which satisfies the Helmholtz equation \eqref{wave-eqn} and
corresponds to an outward-propagating solution \cite{born-wolf-03}.
It is advantageous to perform the angular transformation $\theta=\beta\phi$ 
and to work in the Minkowskian geometry \eqref{eq:metric-Mink} with a wedge
removed rather than in the metric \eqref{eq:metric}, as done in
Ref.~\cite{pla-string16} for an infinitely distant source.
To conveniently perform the transformation, we put the origin at the string
location $S$ and join the point $S$ with the emitting source $E$ by a radial
line [see Fig.~\ref{fig:split}(a)].
Then we assign the values $\phi^-=-\pi$ to the left and $\phi^+=\pi$ to the
right of the line $SE$ that will be the cut line.
Assuming that the emitting wave is symmetric (isotropic), we obtain a zero
derivative $\partial_\phi U=0$ at the cut.
After the angular transformation, the line $SE$ converts to the wedge $SE^-$,
$SE^+$, given by the angles $\pm(\pi-\Delta)$ [see Fig.~\ref{fig:split}(b)].
The two faces of the wedge should be identified since they represent the same
plane in the spacetime \eqref{eq:metric}.
Thus, the propagation of a wave in conical spacetime can be represented as the propagation
of two waves in flat geometry with a wedge removed. In our consideration, each
emitting source lies on the corresponding face of the wedge.
Our next step is to show that the problem posed in this section can be
effectively treated in the framework of the canonical problem of diffraction on a
perfectly conducting half-plane screen \cite{pla-string16}.

\section{Uniform asymptotic theory of diffraction on a half plane}
\label{sec:hp}

\begin{figure}[h!]
\centering
\includegraphics[width=0.5\columnwidth]{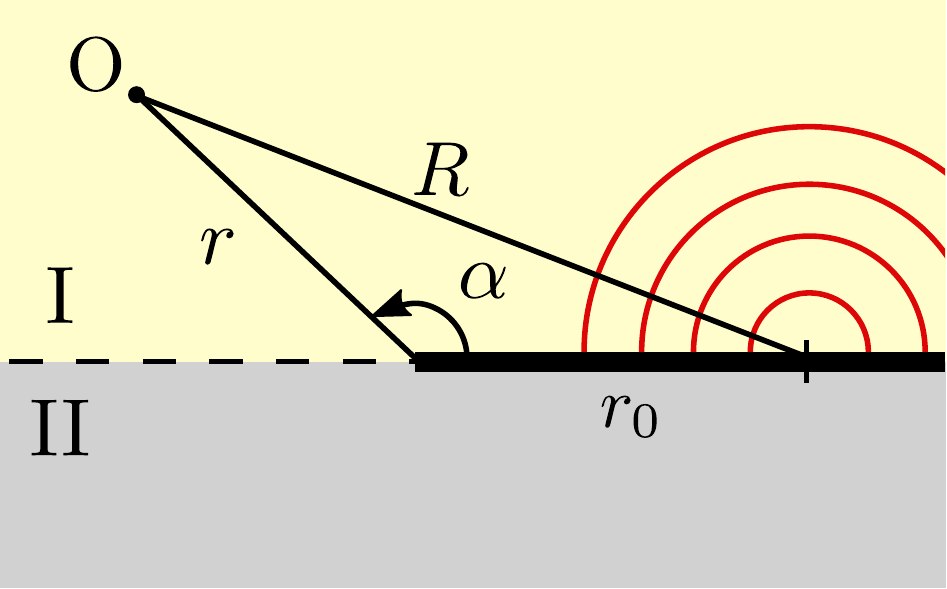}
\caption{Cylindrical wave emitted from a source on the upper surface of a
half-plane screen (thick line). The space is split into two regions:
illuminated (I), shadow (II).}
\label{fig:half-plane-cyl}
\end{figure}

Let us consider a half-plane screen defined in polar coordinates $(r,\alpha)$
by: $\alpha=0$ (upper surface) and $\alpha=2\pi$ (lower surface).
According to our geometry, the source is located on the upper surface of the
screen at a distance $r_0$ from the edge, i.e., at  $(r_0,0)$
(see Fig.~\ref{fig:half-plane-cyl}).

The emission field is a cylindrical wave that can be defined by
\cite{born-wolf-03}
\begin{equation}
U_i= \sqrt{\frac{\pi}{2}} \eee{\ii\pi/4} \,H_0^{(1)} (k R) \approx
\frac{\eee{\ii kR}}{\sqrt{kR}},
\end{equation}
with $R = \sqrt{r^2+r_0^2-2rr_0 \cos \alpha}$ and the subscript ``i'' means
``incident'' field.
The solution for the field in the whole space can be expressed as an integral
\cite{macdonald,clemmow50,bowman69,born-wolf-03}
\begin{equation}
U =  \displaystyle{\sqrt{\frac{2}{\pi}} \, e^{-\ii \pi/4} \, e^{\ii kR}
\int_{-w}^{\infty}
\frac{\eee{\ii \mu^2}\dd \mu}{\sqrt{\mu^2+2kR}} } ,
\label{eq:mac}
\end{equation}
where $w=\sigma\sqrt{k(r+r_0-R})$, with
$\sigma \equiv \mathrm{sgn} \left[\cos \left(\alpha/2 \right)\right] $,
a sign function giving $+1$ in the illuminated region and $-1$ in the shadow
(see Fig.~\ref{fig:half-plane-cyl}).
Instead of working with the solution in the integral representation \eqref{eq:mac},
we apply the uniform asymptotic theory \cite{boersma68,borovikov}, which has
proved to be quite accurate in finding solutions for the wave field under
diffraction. We seek the solution in the form
\begin{equation}
U = U_i \, \mathcal{F}(w) + \mathcal{R},
\label{eq:borov}
\end{equation}
where the first term is the penumbra field with the Fresnel integral defined by
\begin{equation}
\mathcal{F}(u) = \frac{1}{\sqrt{\pi\ii}} \int_{-\infty}^u \eee{\ii \mu^2}\dd \mu
\label{eq:fresnel}
\end{equation}
and $\mathcal{R}$ offsets the residual arising from substituting the penumbra
term in the wave equation.
In both Eqs.~\eqref{eq:mac} and \eqref{eq:borov}, the Neumann boundary condition
for the field is assumed on the screen.
The residual $\mathcal{R}$ has a form of the ray expansion
\begin{equation}
\mathcal{R} = e^{\ii k(r+r_0)}
\sum_{n=1}^{\infty} k^{-n} C_n,
\label{eq:resid}
\end{equation}
in which the slowly varying coefficients $C_n$ can be determined by the method
of asymptotic  matching \cite{borovikov} which consists in comparing the uniform
asymptotics \eqref{eq:borov} with the nonuniform asymptotics of the rigorous
solution and expanding all the terms with respect to inverse powers of $k$.
The nonuniform expansion can be written as
\begin{equation}
U \approx  U_i \,\mathcal{H}(w) +
 U_i^0 \, D\, \frac{\eee{\ii kr}}{\sqrt{kr}}.
\label{eq:asy-no-uni}
\end{equation}
Here, the first term is the geometrical-optics contribution of the incident
wave. It is multiplied by the Heaviside step function $\mathcal{H}(w)$ that
guarantees that this wave only contributes to the illuminated region. 
The second term is the leading order term $\sim O(k^{-1})$ of the diffracted
field. It describes a cylindrical wave emanating from the edge (see, e.g.,
Ref.~\cite{born-wolf-03}). Its amplitude is given by the product of the incident
wave evaluated at the edge, $U_i^0\equiv \eee{\ii kr_0}/\sqrt{kr_0}$, and the
diffraction coefficient \cite{keller62} 
\begin{equation}
D = - \frac{\eee{\ii \pi/4}}{2\sqrt{2\pi}} \frac{1}{\cos(\alpha/2)}.
\label{eq:hp-D}
\end{equation}
Note that the expansion \eqref{eq:asy-no-uni}, valid at $|w|\gg 1$,  is
nonuniform due to a singularity at $w=0$, that is, in the neighbourhood of the
light-shadow boundary.
It should also be remarked that the diffraction coefficient $D$ is determined by
the geometry of the obstacle (a half plane in our case) but is independent of
the type of incident wave \cite{keller62, kouyoumijan74}.

To do the asymptotic matching, the Fresnel integral \eqref{eq:fresnel} is
replaced with its asymptotics at large arguments 
\cite{sommerfeld54}
\begin{equation}
\mathcal{F}(w) \approx \mathcal{H}(w) - \frac{\eee{\ii\pi/4}}{2\sqrt{\pi}}\,
\frac{\eee{\ii w^2}}{w} \equiv  \mathcal{H}(w) -  \mathcal{\tilde{F}}(w).
\label{eq:expansion}
\end{equation}
By comparing Eqs.\ \eqref{eq:borov} and \eqref{eq:asy-no-uni} up to the order
$O(k^{-1})$, we see that only $C_1$ is relevant in the expansion
\eqref{eq:resid} and for the residual $\mathcal{R}$ we obtain
\begin{equation}
\mathcal{R} \approx  U_i \, \mathcal{\tilde{F}} +
 U_i^0 \, D\, \frac{\eee{\ii kr}}{\sqrt{kr}}.
\end{equation}
Substituting in Eq.\ \eqref{eq:borov}, one can write the final solution
\begin{equation}
U \approx  U_i \, (\mathcal{F} + \mathcal{\tilde{F}} ) +
 U_i^0 \, D\, \frac{\eee{\ii kr}}{\sqrt{kr}}.
\label{eq:hp-uni-asy}
\end{equation}
Written in this form, the solution \eqref{eq:hp-uni-asy} corresponds to the
uniform asymptotic theory introduced in Ref.~\cite{boersma68}.
It is called uniform since the poles in the diffraction coefficient $D$ are
cancelled out by the poles in the term $\mathcal{\tilde{F}}$, giving a regular
solution in the whole space including the light-shadow boundary.
It would be convenient to combine both singular terms in one by defining a new
diffraction coefficient
\begin{equation}
\tilde{D} = - \frac{\eee{\ii \pi/4}}{2\sqrt{2\pi}}
\left[  \frac{1}{\cos(\alpha/2)} - \sigma \sqrt{\frac{2rr_0}{R(r+r_0-R)}}
\right],
\label{eq:D-modif}
\end{equation}
and the uniform solution finally becomes
\begin{equation}
U \approx  U_i \, \mathcal{F}(w)
+  U_i^0 \, \tilde{D}\, \frac{\eee{\ii kr}}{\sqrt{kr}}.
\label{eq:hp-uni-asy2}
\end{equation}
Note that Eq.~\eqref{eq:hp-uni-asy2} is valid at any distances from the
light-shadow boundary, near and away from the edge, i.e.~everywhere except for
the neighbourhood of the source, since $kR \gg 1 $ is assumed.
Far from the light-shadow boundary, both asymptotics, the nonuniform
\eqref{eq:asy-no-uni}  and uniform one \eqref{eq:hp-uni-asy2}, coincide.
We also observe that the edge wave has a phase shift of $-3\pi/4$ in the
illuminated region and $+\pi/4$ in the shadow with respect to the incident
field. 
Crossing the shadow line introduces a phase change of $\pi$, which is
manifested in the sign change of the diffraction coefficients $D$ and
$\tilde{D}$
(See the original work by Fresnel \cite{fresnel1821,crew1900} who pointed out that the diffracted waves
in the shadow and illuminated regions are in complete phase opposition).

\section{Diffraction of a cylindrical wave by a cosmic string}

As explained in Sect.~\ref{sect:wave-eq}, after an angular transformation, the 
spacetime \eqref{eq:metric} can be represented in the flat Minkowskian geometry 
\eqref{eq:metric-Mink} with a wedge of $2\Delta$ removed [Fig.~\ref{fig:split}(b)]. 
Accordingly, the wave
source $E$ is doubled into images $E^-$, $E^+$ which are located on the faces of
the wedge. Each image source emits a cylindrical wave that will be diffracted by
the corresponding half plane.
Therefore, the wave diffraction by a string can be thought of as the
diffraction by two half planes forming an angle of $2\Delta$ \cite{pla-string16}.

\begin{figure}[h!]
\centering
\includegraphics[width=0.95\columnwidth]{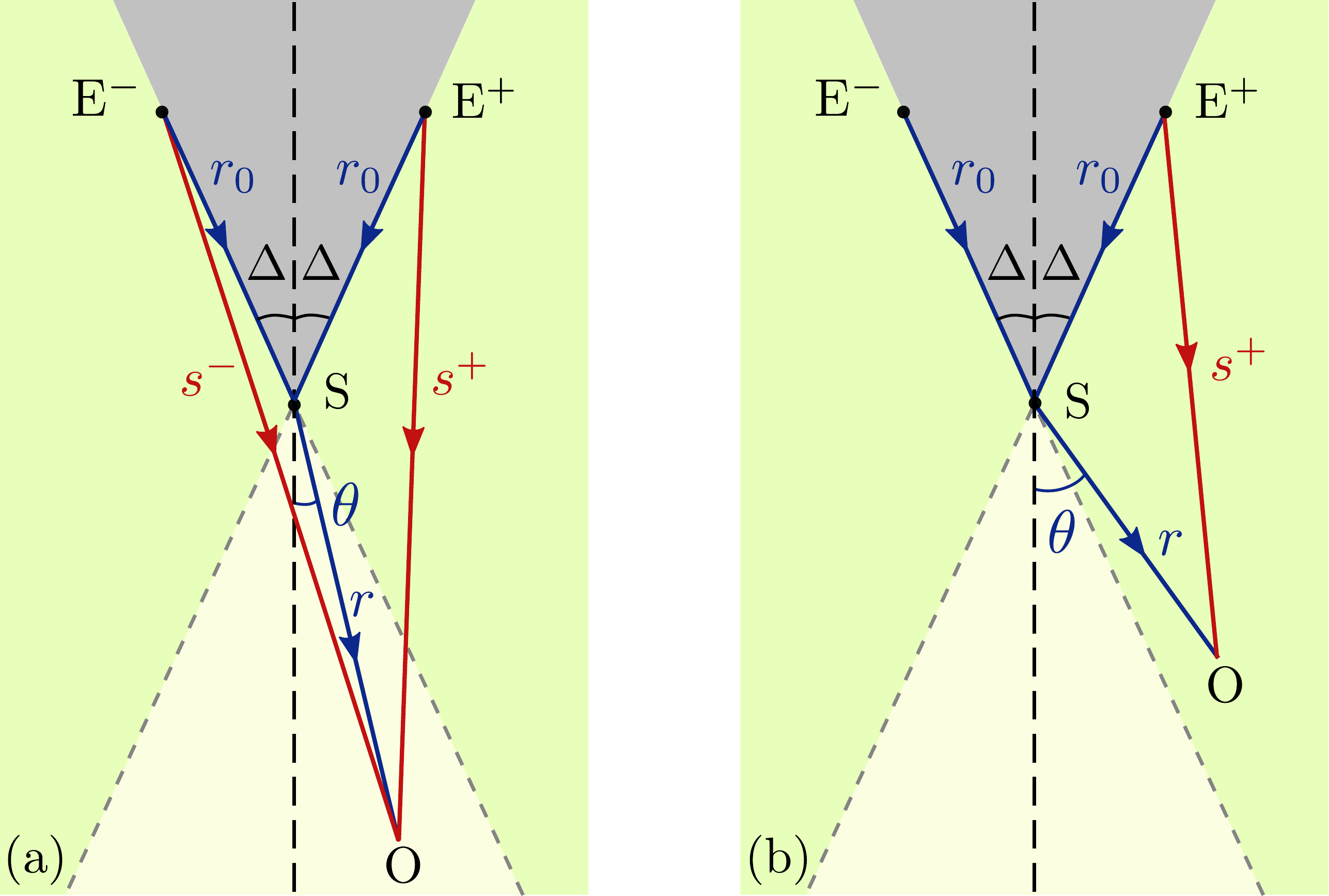}
\caption{Geometrical theory of diffraction in conical space.
Direct (red) and diffracted (blue) rays determine the leading order contribution
at the observation point $O$, when $O$ is either in the double-imaging (a)
or single-imaging region (b).}
\label{fig:geom}
\end{figure}

We now construct the wave field by making use of the results described in the
previous section for the case of a single half plane.
 We use the angular substitution $\alpha=\pi-\Delta\mp\theta$ for each half plane in order to work with the flat coordinates $(r,\theta)$.
The emission field for each source is described by a cylindrical wave given by
\begin{equation}
U_i^\pm \approx \frac{\eee{\ii ks^\pm}}{\sqrt{ks^\pm}},
\end{equation}
where $s^\pm=\sqrt{r^2+r_0^2+2rr_0\cos(\Delta\pm\theta)}$.
From Eq.~\eqref{eq:hp-uni-asy2}, the uniform asymptotic solution for the field
at the observation point $(r,\theta)$ is found in the form
\begin{equation}
U=U_i^-\mathcal{F}(w^-)+U_i^+\mathcal{F}(w^+)
+U_i^0 \left(\tilde{D}^- +\tilde{D}^+ \right) \frac{\eee{\ii kr}}{\sqrt{kr}},
\label{eq:s-uni-asy}
\end{equation}
with the modified diffraction coefficients $\tilde{D}^\pm$ defined as
\begin{equation}
\tilde{D}^\pm=- \frac{\eee{\ii \pi/4}}{2\sqrt{2\pi}}
 \left[  \frac{1}{ \sin{[\frac{1}{2}(\Delta\pm\theta})] }
- \sigma^\pm \sqrt{\frac{2rr_0}{s^\pm(r+r_0-s^\pm)}} \right],
\label{eq:diff-pm}
\end{equation}
and the notations
$w^\pm=\sigma^\pm\sqrt{k(r+r_0-s^\pm)}$ and $\sigma^\pm=\text{sgn}(\Delta\pm\theta)$.
According to the values of the sign functions $\sigma^\pm$, the entire space
(beyond the wedge) is divided into several regions of interest (see
Fig.~\ref{fig:geom}): (i) a double-imaging region, $-\Delta<\theta<\Delta$,
illuminated by both sources and (ii) two single-imaging regions illuminated by just
one image source (compare with similar geometry of Ref.~\cite{pla-string16}).

Far from the boundaries, $\theta = \pm\Delta$, one can use the geometrical
theory of diffraction \cite{keller62} which corresponds to the nonuniform
expansion \eqref{eq:asy-no-uni} for a half-plane solution.
In our case of a double source, we obtain
\begin{equation}
U = U_i^- \mathcal{H}(w^-) + U_i^+ \mathcal{H}(w^+)  + U_i^0
\left( D^- + D^+ \right)
\frac{\eee{\ii kr}}{\sqrt{kr}}.
\label{eq:s-non-asy}
\end{equation}
The first two terms with the Heaviside functions describe the geometrical optics
(GO) waves. The step functions guarantee that the GO waves only contribute to
the respective illuminated regions (Fig.~\ref{fig:geom}).
The third term is the leading order term of the diffracted (D) field. 
It describes a cylindrical wave emanating from the edge and whose amplitude
depends on the diffraction coefficients (which can also be called ``directivity
functions'') determined by
\begin{equation}
D^\pm = - \frac{\eee{\ii \pi/4}}{2\sqrt{2\pi}}
\frac{1}{\sin{[\frac{1}{2}(\Delta\pm\theta})]}.
\label{eq:diff-D}
\end{equation}
Note that the D wave has a phase shift of $3\pi/4$ whenever the observation
point is in the double-imaging region.
The terms in Eq.~\eqref{eq:s-non-asy} are visualised in Fig.~\ref{fig:geom},
where each contribution corresponds to a characteristic ray: two GO rays
going from the sources $E^\pm$ to the observer $O$ directly and two D rays going
from the sources but hitting the edge $S$ -- the string location -- following
the shortest path (Fermat's principle for edge diffraction \cite{keller62}).

It is easy to check that for $\Delta=0$, i.e. when there is no lensing due to
string, both Eqs.~\eqref{eq:s-uni-asy} and \eqref{eq:s-non-asy} reduce to the
unlensed field 
\begin{equation}
U_0=\frac{\eee{\ii ks_0}}{\sqrt{ks_0}},
\end{equation}
which is a usual cylindrical wave with $s_0=\sqrt{r^2+r_0^2+2rr_0\cos\theta}$.
For future analysis, one can define the amplification factor $F=U/U_0$ to
characterize the effect of gravitational lensing by the string over the wave field.
Finally, it can be verified that one recovers all the expressions derived for
the plane wave in Ref.~\cite{pla-string16} by multiplying the line-source results
by $\sqrt{kr_0}\eee{-\ii k r_0}$ and letting $r_0\to\infty$ \cite{clemmow50}.

\section{Fresnel observation zones}

The Fresnel-zone concept has been widely used in various branches of wave physics.
When the wave field is calculated at a certain observation point, it is
advantageous to divide an incoming wavefront into a number of zones,
each with an additional path difference of a half-wavelength, so that the
wavefront phase changes by $\pi$ when moving from one zone to the next
\cite{sommerfeld54,born-wolf-03}.
The construction of these zones provides a pictorial understanding of the
diffraction phenomenon. Indeed, when some of the zones are obstructed by a
screen or any other obstacle, the Fresnel zones are used to determine
qualitatively when the diffraction effects become important and the geometrical
optics limit is not accurate to estimate the resulting field.
For example, when radio waves propagate in terrain environment, Fresnel
zones are elliptic-shaped regions surrounding the line-of-sight path from source
to receiver \cite{bertoni99}.
To achieve an acceptable transmission not affected by diffraction or multipath
attenuation, all disturbing objects must be further than 0.6 times the first Fresnel
zone radius from the line-of-sight path \cite{bertoni99}.

It should be noted that diffraction may appear in the absence of screens or
obstacles which might obstruct the direct wave transmission. As Fresnel pointed
out in his classical work \cite{fresnel1821,crew1900}, in order to produce the
phenomena of diffraction ``all that is required is that a part of the wave
should be retarded with respect to its neighbouring parts.''
This is precisely what happens in the lensing effect.
When a gravitational or electromagnetic wave passes near a massive cosmological
object, it deviates giving rise to multiple images and diffraction
\cite{deguchi86,schneider92,nakamura99}.
A similar effect may appear when the wave propagates near a topological defect
like a cosmic string \cite{suyama06,pla-string16,yoo13}.

\subsection{Hyperbolic vs elliptic zones}

\begin{figure}[t]
\centering
\includegraphics[width=0.95\columnwidth]{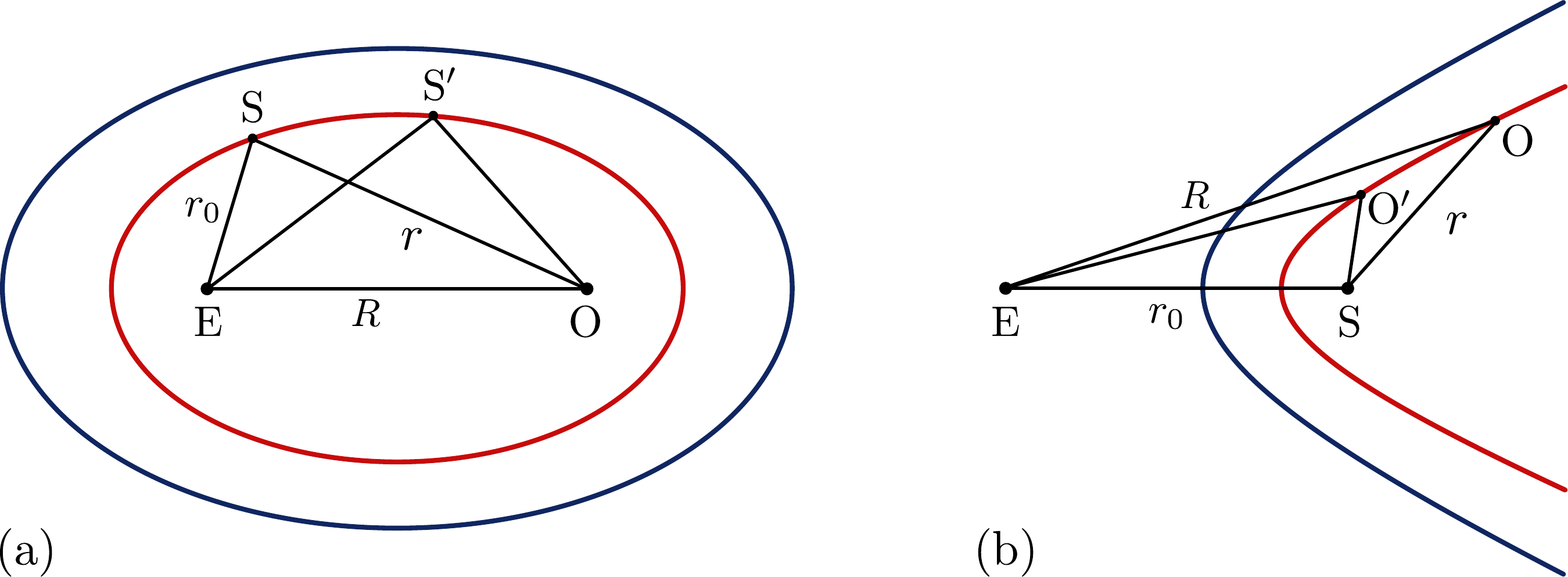}
\caption{
Lines of constant phase difference between the direct path $|EO|=R$ and the
diffracted path $|ES|+|SO|=r_0+r$. Their shape is: (a) elliptic when the
distance $R$ is  fixed; (b) hyperbolic when $r_0$ is fixed.}
\label{fig:elliptic-hyper}
\end{figure}

For the problem of transmission of a signal from the emitter $E$ to the observer
$O$, the Fresnel zones can be constructed about the line of sight $EO$
connecting the points. In this case, the relevant geometry
is a set of confocal elliptic-shaped regions with the foci at the points
$E$ and $O$ \cite{bertoni99}.
Indeed, when any obstacle $S$ is not far from the line of sight, an alternative
path $ES+SO$ interferes with the direct path $EO$ resulting in constructive or
destructive interference.
The result depends on the phase difference between the paths
[see Fig.~\ref{fig:elliptic-hyper}(a)]. 
If the points $E$ and $O$ are fixed, while $S$ is moved over the space, the line
of constant phase difference is elliptic.
The objective of such construction is to determine clearance zones in order to
achieve perfect transmission between the source and receiver \cite{bertoni99}.
However, for our case this zone construction is not convenient since (i) we
have two image sources instead of one, and (ii) we are interested in just the
opposite --- in how to detect the scattering object due to the presence of wave
effects in the observed signal, or in other words, where we should place the
observer with the aim to detect the obstacle by virtue of diffraction with the
highest efficiency.
To this purpose we fix the points $E$ and $S$, while the observer $O$ is moved
over the space [Fig.~\ref{fig:elliptic-hyper}(b)]. By this procedure the line of
constant phase difference between the two paths will be hyperbolic instead of
elliptic (at the moment we assume a Minkowskian geometry).

\subsection{Half plane}
\begin{figure}[t]
\centering
\includegraphics[width=\columnwidth]{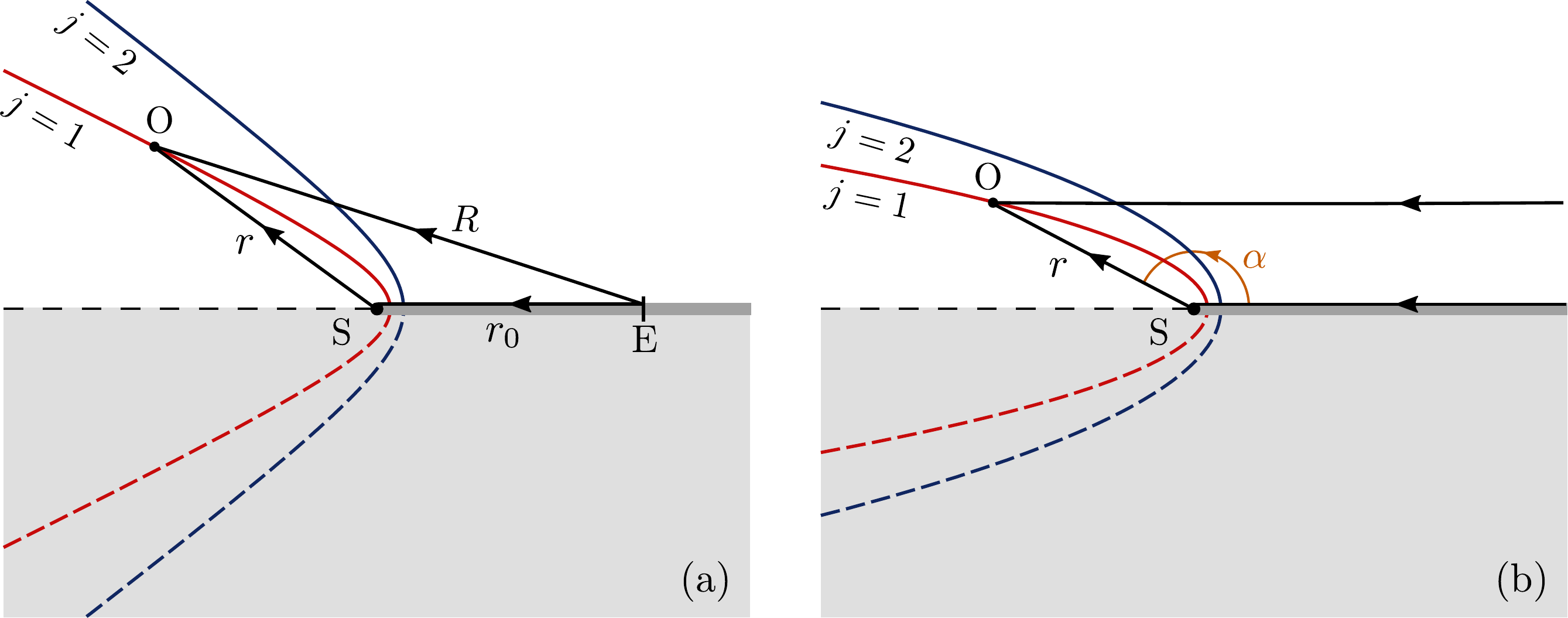}
\caption{Lines of constant phase for the case of diffraction on a
half-plane screen: 
(a) the source $E$ is at a distance $r_0$ from the edge $S$, the lines are hyperbolic;
(b) the source is at infinity (plane wave incidence), they are parabolic.}
\label{fig:hyp-hp}
\end{figure}

This idea can easily be implemented to our equations.
First, consider the case of a single half plane with the geometry of
Fig.~\ref{fig:half-plane-cyl}. The penumbra term in Eq.~\eqref{eq:hp-uni-asy2}
is determined by the Fresnel function of argument $w$, that depends on the path
difference $d \equiv r+r_0-R$.
Hence, one can define the zone structure by the condition that $d$ should be an
integer number of half wavelengths:
\begin{equation}
r+r_0 - \sqrt{r^2+r_0^2-2rr_0 \cos \alpha}
 = \frac{\lambda}{2} \,j,
\label{eq:fres-zone}
\end{equation}
that means the phase between the paths changes by $\pi$ when moving from one
zone to the next -- a similar argument used by Fresnel to define the zone
boundaries on the wavefront \cite{fresnel1821,crew1900}.
From this equation, the shape of the zones in polar coordinates $(r,\alpha)$
will be determined by a familiar expression for conic sections
\begin{equation}
r=\frac{r_0}{2e_j}\, \frac{e_j^2-1}{1 + e_j \cos \alpha}
\label{eq:hyper}
\end{equation}
with eccentricity $e_j$ given by
\begin{equation}
e_j \equiv \left(1-\frac{\lambda}{2r_0}\,j \right)^{-1}.
\label{eq:e}
\end{equation}
Eq.~\eqref{eq:hyper} describes a set of confocal hyperbolas (for
$j=1,2,3,\dots$), all with the foci at the source $E$ and the edge of the
screen $S$,  so that the path difference from each focus to a point on the curve
is constant [see Fig.~\ref{fig:hyp-hp}(a)].
Each hyperbola, in principle, should have two symmetrical branches corresponding
to either $R>r$ or $R<r$. Since the latter case is not relevant for our physical
conditions, we only consider the branches in the neighbourhood of the focus $S$
with $R>r$. 
Moreover, we are only interested in those parts of the branches which are
located in the upper (illuminated) region where the direct and diffracted paths
interfere.
In the lower (shadowed) region no interference effects are expected
since there is only the diffracted wave (this part is plotted in a dashed line).
The vertices of the hyperbolas are given by the coordinates
($\frac{1}{4}\lambda j, 0$), so that they are equidistantly spaced by
$\frac{1}{4}\lambda$ between $r=0$ and $r=r_0/2$ along the screen.
The asymptotes of the hyperbolas determine the limits for the angle $\alpha$ for
each zone: $ -\alpha_j^*<\alpha< \alpha_j^*$ with $\alpha_j^*=\arccos
[-e_j^{-1}]$.
We also notice that for our geometry, due to the requirement $1<e_j<\infty$ for
hyperbolas, there is an upper limit for the index $j$ given by 
$j_{\rm max}=2r_0/\lambda$.
Finally, for an infinitely distant source (incident plane wave), one focus goes
to infinity, $r_0\to\infty$, and the path difference is simply
$d = r (1+\cos \alpha)$, while the hyperbolas become parabolas with the single
focus at the edge $S$ [see Fig.~\ref{fig:hyp-hp}(b)].
The shape of the parabolas is determined by
\cite{pla-string16}
\begin{equation}
r=\frac{\lambda}{2} \,\frac{j}{1+\cos \alpha}.
\label{eq:parab}
\end{equation}

\subsection{String}
\label{fresnel-string}
Next, we consider the geometry of Fig.~\ref{fig:split}(b) corresponding to the
string, which is flat space with a wedge removed and two sources located on
the faces of the wedge.
A qualitative analysis of Eq.~\eqref{eq:s-non-asy} shows that in the most 
interesting situation, when the observer is in the double-imaging region, the
diffraction pattern will be determined by the interference of four characteristic
waves: two GO waves coming from the sources and two D waves emanating from the
edge [Fig.~\ref{fig:geom}(a)].

First of all, from geometrical optics we would expect the following picture:
two GO waves interfering with each other, constructively or destructively, to
produce an interference pattern of bright and dark lines alternating in space.
The phase difference between the GO waves is constant along the lines:
$s^- - s^+ = const$, which are confocal hyperbolas with the foci at
the sources $E^-$ and $E^+$.
If we specify the path difference in units of the half wavelength:
\begin{equation}
 s^- - s^+ = \frac{\lambda}{2} \, q
\label{eq:go-interf}
\end{equation}
with $q \in\mathbb{Z}$ being an integer, the bright lines (constructive
interference) correspond to even $q=0, \pm 2, \pm 4, \dots$, while the dark
lines (destructive interference) correspond to odd values  $q=\pm 1, \pm 3,
\dots$. In the following, we will refer to the bright and dark GO lines as
``antinodal'' and ``nodal'' lines, respectively.

\begin{figure*}[t!]
\centering
\includegraphics[width=0.8\columnwidth]{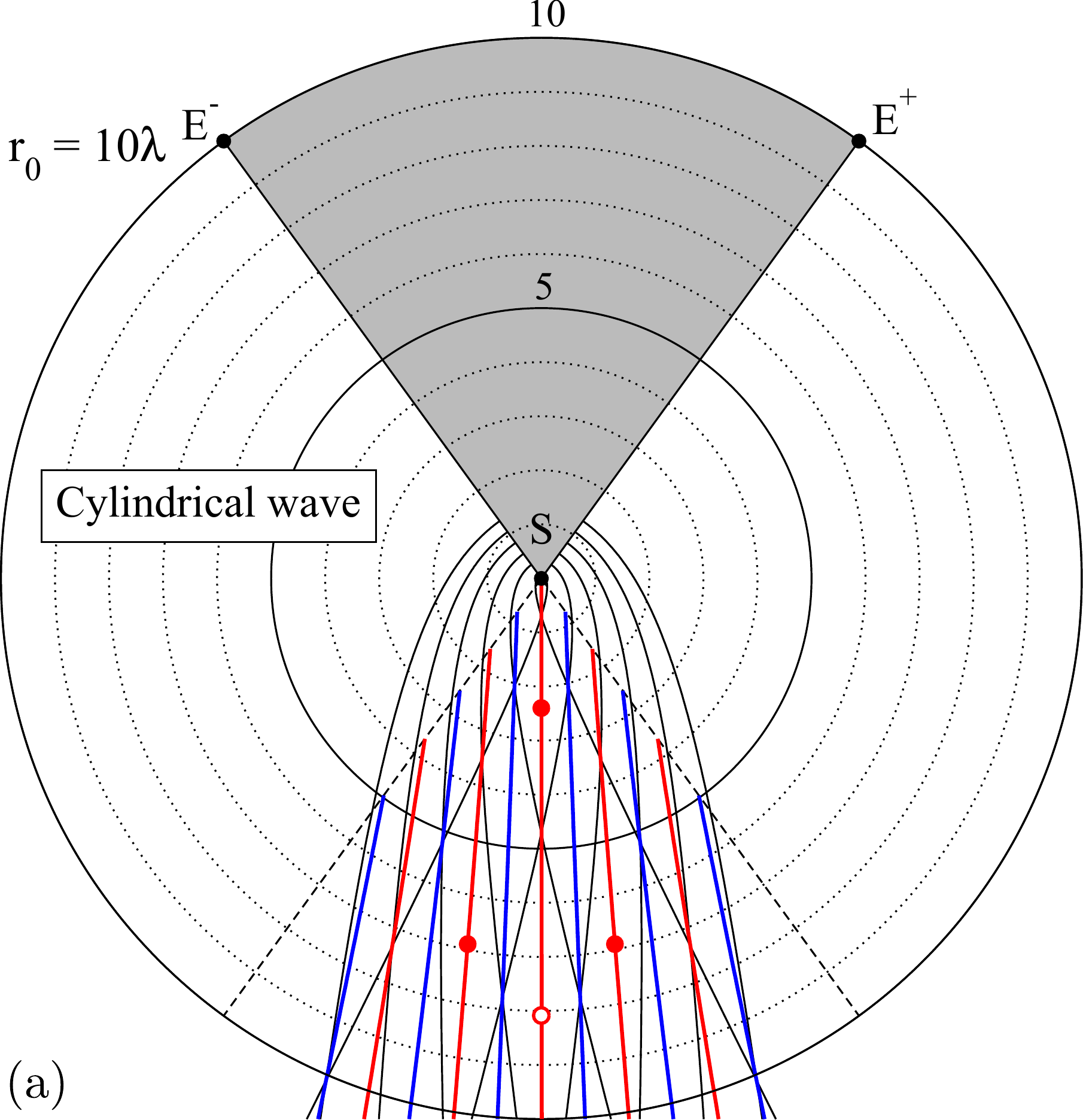}
\includegraphics[width=0.8\columnwidth]{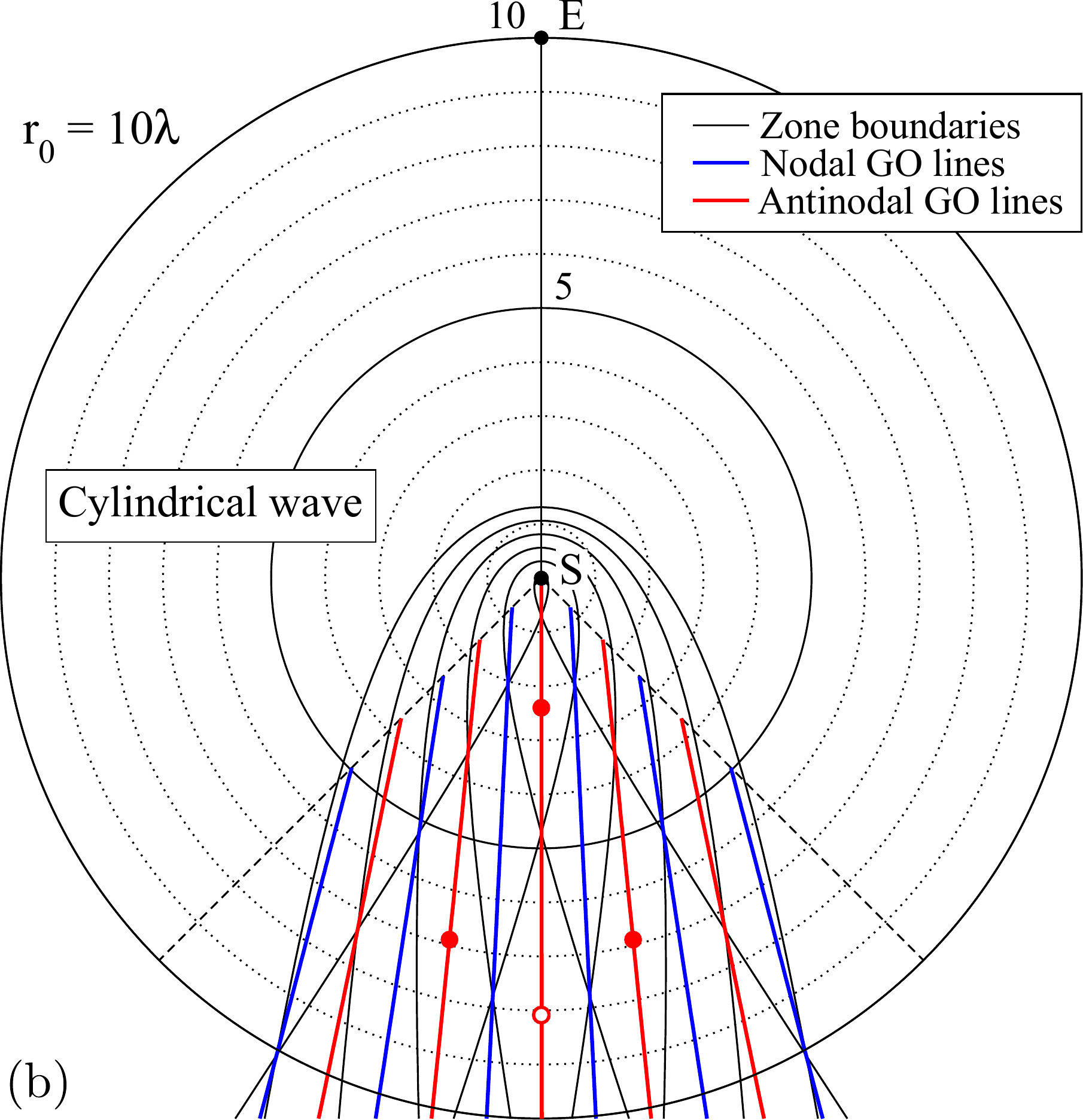}\\
\vspace*{2ex}
\includegraphics[width=0.8\columnwidth]{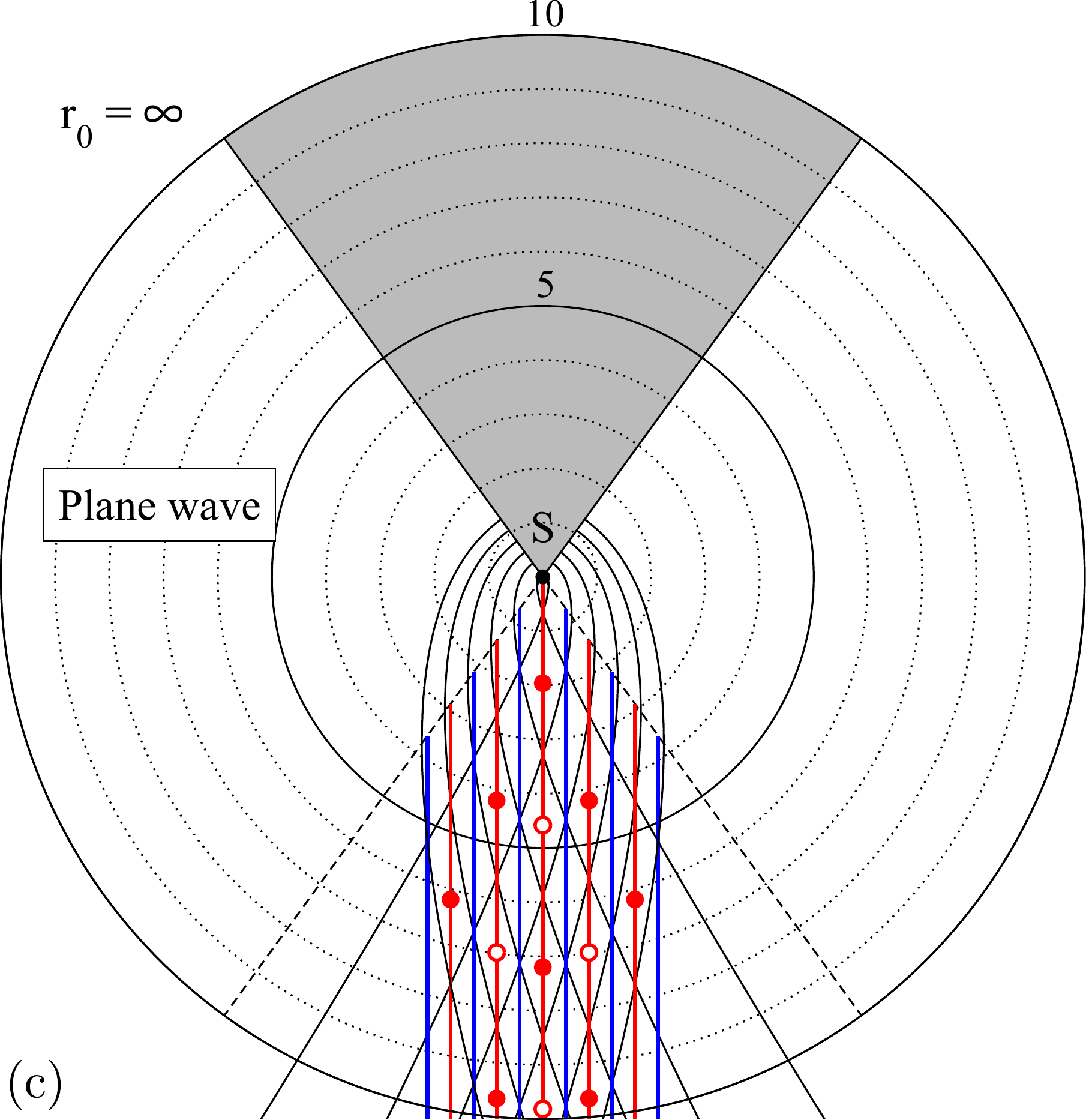}
\includegraphics[width=0.8\columnwidth]{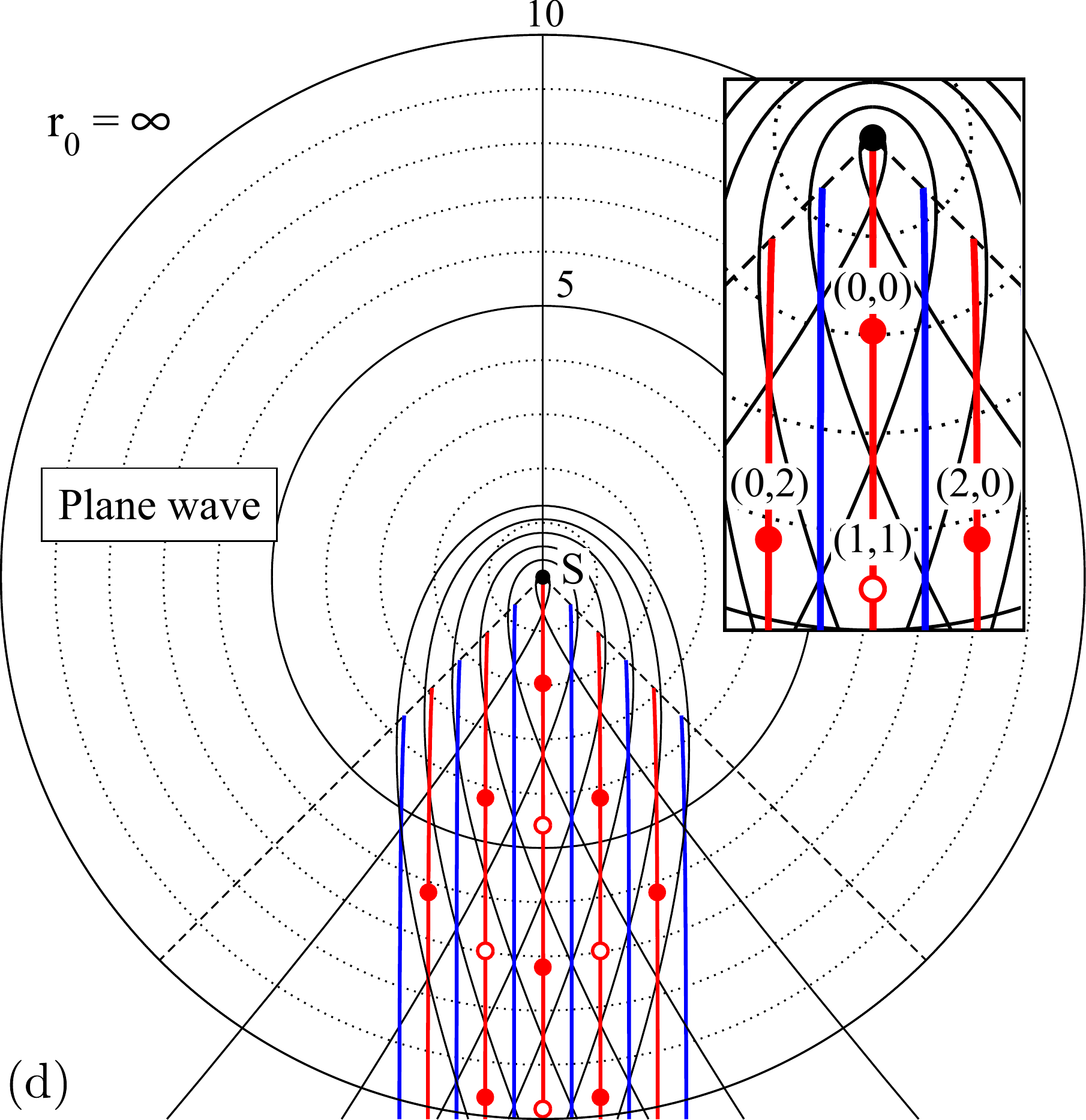}
\caption{Observation zones associated with interference and diffraction of waves
by a cosmic string:
(a),(b) the source $E$ is at a finite distance $r_0$ from the string $S$;
(c),(d) the source is at infinity (incident plane wave).
Each case is shown for two coordinate systems: 
(a),(c) Minkowskian \eqref{eq:metric-Mink} with a wedge removed,
and (b),(d) the one corresponding to Eq.~\eqref{eq:metric}.
For a better view, a rather large value
$\Delta=0.2\pi$ is taken. Dashed lines indicate the limits of the double-imaging
region.
Red points are the maxima of the field intensity, while red circles correspond
to saddle points. Inset in (d): the points are labelled with $(n,m)$ indices.}
\label{fig:hyp}
\end{figure*}

The diffracted waves introduce new important features into the overall
interference pattern.
As we pointed out for the case of a half plane, the phase difference between the
GO and D waves is constant along the hyperbolas:
$r+r_0 - s^- = const$, $r+r_0 - s^+ = const$,
for the sources $E^-$ and $E^+$, respectively.
The condition for destructive and constructive interference will now be
different from that of Eq.~\eqref{eq:go-interf}.
D waves acquire an additional phase shift of $3\pi/4$ by hitting the edge, which
is manifested by virtue of the phase in the diffraction coefficients
\eqref{eq:diff-D}. Therefore, we would expect the maxima and minima
of the field intensity when these two conditions are fulfilled simultaneously:
\begin{align}
\label{eq:max-min1}
r+r_0 - s^+  &=  \frac{\lambda}{2} \, \left(n+\frac{3}{4} \right), \\
r+r_0 - s^-   &=  \frac{\lambda}{2} \, \left(m+\frac{3}{4} \right)
\label{eq:max-min2}
\end{align}
with $m,n$ being non-negative integers: $0, 1, 2, 3, \dots$.
The solutions are the intersection points of two families of hyperbolas
corresponding to each source. If we now subtract Eqs.~\eqref{eq:max-min1} and
\eqref{eq:max-min2}, we get Eq.~\eqref{eq:go-interf} with $q=n-m$, that means
these intersection points lie precisely on the nodal and antinodal GO lines. 
Therefore, we would expect that the additional interference with the D waves may
lead to a further amplification of the field on the antinodal lines.
The points of considerable interest are the global maxima, which occur when the two
GO and the two D waves are all in phase,
that corresponds to having all three numbers, $n$, $m$, and $q$, even.
Denoting the intersection points by a pair of numbers $(n,m)$, the highest
maximum occurs at the point $(0,0)$ which is located at the line of sight
(central antinodal line).
The next-order maxima are $(0,2)$ and $(2,0)$ lying symmetrically out of the
line of sight at a larger distance from the string and having, therefore, lower
magnitude. They are followed by more distant maxima $(0,4)$, $(2,2)$, $(4,0)$,
and so on (see Fig.~\ref{fig:hyp}).
An important special case occurs when $n$ and $m$ are odd simultaneously
(accordingly $q$ is even). These are the saddle points of the field intensity
which are located at the antinodal GO lines, e.g., $(1,1)$, $(3,1)$, $(1,3)$,
etc.
On the other hand, on the nodal lines, D waves do not substantially affect the
wave field intensity due to the destructive interference between the GO waves.

For further analysis, one can define the observation zones, $\mathcal{Z}_{nm}$,
associated with the points $(n,m)$, which are characteristic points of
interference between the GO and D waves.
Since an increase by 1 in indices corresponds to changes by $\pi$ in phase,
we define the zones as delimited by the hyperbolic lines \eqref{eq:max-min1} and
\eqref{eq:max-min2} with the substitution:
$n\rightarrow n\pm \frac{1}{2}$ and $m\rightarrow m\pm \frac{1}{2}$.
We call $\mathcal{Z}_{nm}$ as ``Fresnel observation zones'', since the zone
structure is basically determined by Fresnel diffraction.
The zone boundaries can be defined explicitly by the hyperbolas (see
Eq.~\eqref{eq:hyper} for comparison):
\begin{equation}
r=\frac{r_0}{2e_j}\, \frac{e_j^2-1}{1 - e_j \cos(\Delta\pm\theta)}
\label{eq:hyper2}
\end{equation}
with eccentricity 
\begin{equation}
e_j = \left[1-\frac{\lambda}{2r_0}\, \left(j+\frac{3}{4} \right) \right]^{-1}.
\label{eq:e2}
\end{equation}
For each zone $\mathcal{Z}_{nm}$ we have to substitute:
$j  = n - \frac{1}{2}; \, n + \frac{1}{2}$ for the upper sign,
$j  = m - \frac{1}{2}; \, m + \frac{1}{2}$ for the lower sign in
Eq.~\eqref{eq:hyper2}.
(Two different signs refer to the image sources $E^\pm$).
The structure of Fresnel observation zones is depicted in Fig.~\ref{fig:hyp}.
Here, the hyperbolas \eqref{eq:hyper2} are shown in black along with the GO
antinodal (in red) and nodal (in blue) lines.
Note that the interference between the GO waves takes place only in the
double-imaging sector (bounded by dashes). Outside of it, the wave field is
determined by interference of one GO and two D waves. Therefore, one would
expect in the single-imaging region, the bright (antinodal) and dark (nodal)
lines to be given by Eq.~\eqref{eq:max-min1} to the right, and
Eq.~\eqref{eq:max-min2} to the left of the string location.

For an infinitely distant source, $r_0 \to \infty$, the plane-wave approximation
for the incident wave is held.
Fig.~\ref{fig:hyp}(c),(d) shows the corresponding Fresnel-zone structure. 
In this case, the antinodal and nodal GO lines of Eq.~\eqref{eq:go-interf}
simply become straight lines parallel to the line of sight and given by
\begin{equation}
2 r \sin\theta \,\sin\Delta = \frac{\lambda}{2} \, q
\label{eq:go-interf2}
\end{equation}
with $q$ being an integer.
One can see that the separation between these lines is constant.
This means that the typical fringe separation in the observation plane is
$\lambda/(2\,\sin\Delta) \approx\lambda/(2\Delta)$, which is independent of the
distance (even in space \eqref{eq:metric}  whenever $\Delta\ll 1$).
This is different from what happens in the diffraction by a compact object
\cite{nakamura98}.
For plane waves, the lines of constant phase between the GO and D waves become
parabolas \cite{pla-string16}: $r[1-\cos(\Delta\pm\theta)] = const$.
On the other hand,  the phase shift in diffraction coefficients does not change
when the source goes to infinity, therefore the conditions to find the maxima
and minima will be similar to Eqs.~\eqref{eq:max-min1} and \eqref{eq:max-min2},
in which the indices $n$ and $m$ will identify the intersections of the
parabolas. Since parabolas are the conic sections with eccentricity $e=1$, the
parabolic Fresnel zones will be determined simply by
\begin{equation}
r=\frac{\lambda}{2} \,\frac{j+3/4}{1-\cos(\Delta\pm\theta)}
\label{eq:parab2}
\end{equation}
with $j$ defined below Eq.~\eqref{eq:e2}.
These zones are depicted in Fig.~\ref{fig:hyp}(c).
Finally, in order to obtain the observation zones in space \eqref{eq:metric}, the angular
transformation $\theta=\beta\phi$ should be performed.
As shown in Fig.~\ref{fig:hyp}(b),(d), this angular stretching distorts somewhat
the shape of the curves, particularly as the angle increases.
On the string's line of sight ($\theta$=$\phi$=$0$), however, the boundaries
between the zones, as well as the maxima, coincide for both backgrounds (see
Fig.~\ref{fig:hyp}).

The construction of Fresnel zones can also be carried out for other geometries.
For instance, one can study a three-dimensional case with a point source
emitting spherical waves, for which the analytical formulas for diffraction on a
half plane are also known \cite{macdonald,bowman69}.
The surfaces of constant phase between the GO and D waves, though with more
involved shapes, can also be obtained.
To find the global diffraction maxima, what is needed is the value of the phase
shift acquired by the D wave when hitting the string following the shortest
path. It does not depend on the type of incident wave but on the obstacle
\cite{keller62, kouyoumijan74}, having the value of $3\pi/4$ we have found for
the conical space \eqref{eq:metric}.
We also note that for a far distant source, one can neglect the curvature of
the wavefront and use the plane-wave approximation.
In this limit, we expect the Fresnel zones to be hyperbolic cylinders having one
focus on the string and the position of the other will depend on the tilted
angle of incidence. In case of perpendicular incidence, the cylinders will
become parabolic.

\section{Discussion of the results}

The zone structure we have introduced by simple analysis of four-wave
interference is based on the geometrical theory of diffraction prescribed by
Eq.~\eqref{eq:s-non-asy}.
In spite of its asymptotic character, this theory is known to fit almost
perfectly the exact solution in the diffraction experiments on obstacles as
small as two wavelengths, with good predictions down to one wavelength
[see, e.g., Ref.~\cite{kapany65}].
\begin{figure}[!tb]
\centering
\includegraphics[width=0.9\columnwidth]{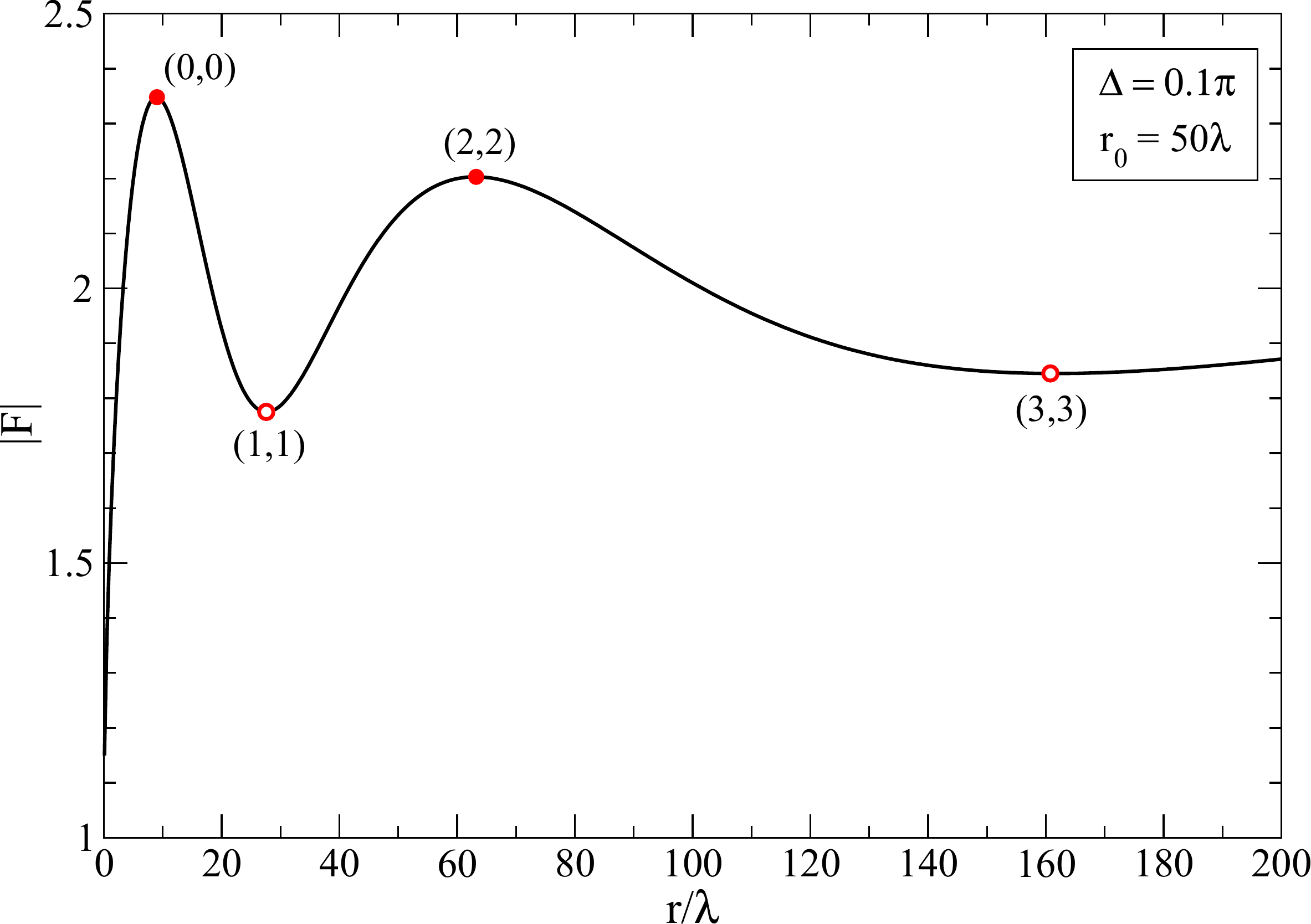}
\caption{Amplification factor vs distance $r$ when the string is on the line of
sight (central antinodal line). Intersection points of the hyperbolas are
labelled.}
\label{fig:centr-antinod}
\end{figure}
\begin{figure}[!tb]
\centering
\includegraphics[width=0.85\columnwidth]{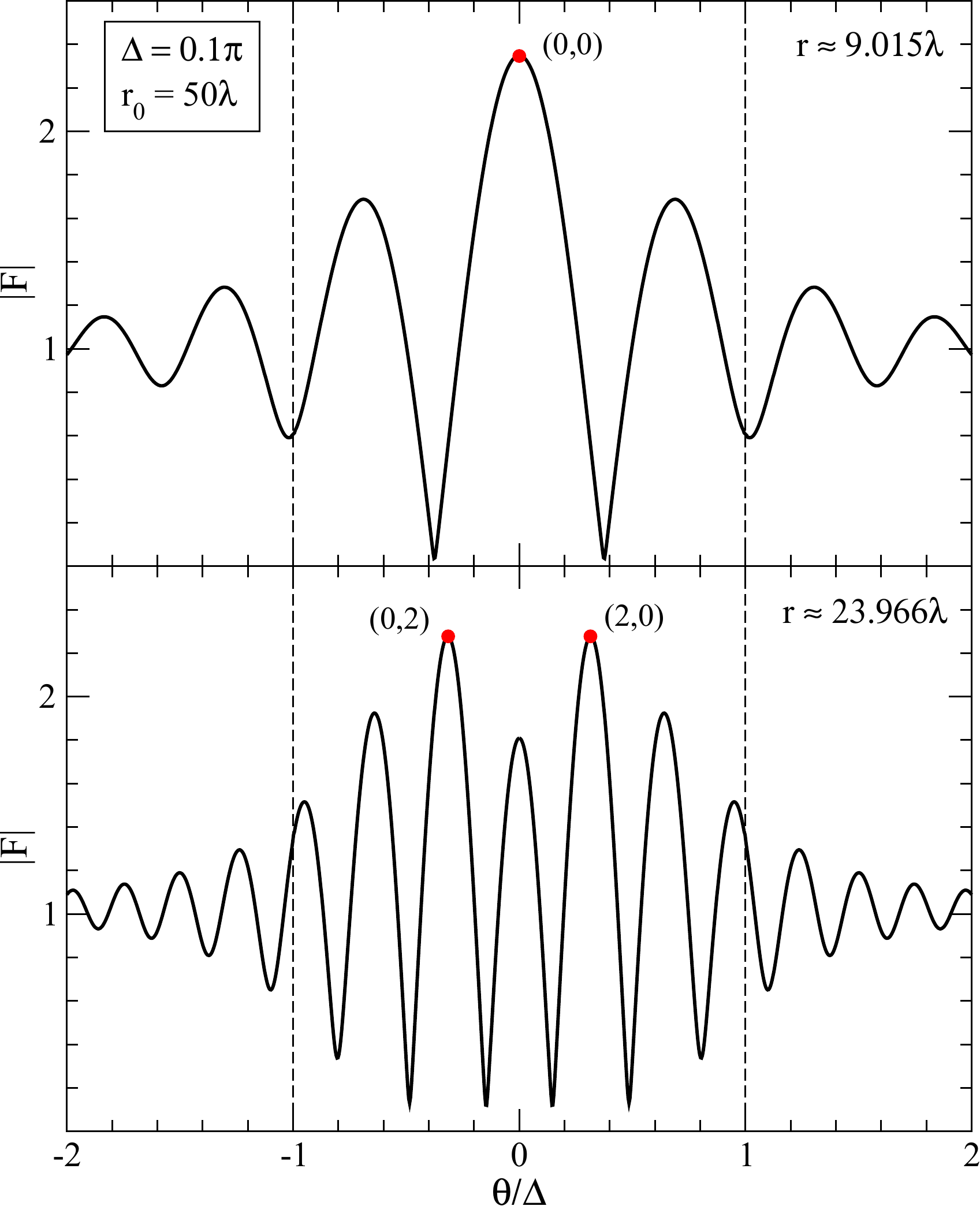}
\caption{Amplification factor vs angular coordinate $\theta$ normalized to
$\Delta$ for fixed distances from the string:  (a) $r\approx9.015\lambda$
corresponding to the position of $(0,0)$; (b) $r\approx23.966\lambda$
corresponding to the location of $(2,0)$ and $(0,2)$.
The boundaries of the double-imaging region are indicated by dashes.}
\label{fig:transv}
\end{figure}
We therefore believe that one can predict the location of the diffraction maxima
due to wave scattering on the cosmic string with very high accuracy by a simple
procedure described in Sect.~\ref{fresnel-string}.
For instance, to find the maximum $(n,m)$ corresponding to the zone
$\mathcal{Z}_{nm}$, all we need is to calculate the crossing point of the two
hyperbolas given by Eq.~\eqref{eq:hyper2} with $j=n$ and $j=m$ for the two
sources, respectively. This determines at which distance $r$ and angle $\theta$
from the string the point of maximum intensity should appear.
To confirm our finding, Figs.~\ref{fig:centr-antinod} and \ref{fig:transv} plot
the modulus of the amplification factor $F$=$U/U_0$ calculated from the uniform
asymptotics \eqref{eq:s-uni-asy}, which is more accurate than the geometrical
theory of diffraction. The case when the string is on the line of sight, that
corresponds to the observer on the central antinodal line ($\theta=0$), is
depicted in Fig.~\ref{fig:centr-antinod}.
The intersection points of the hyperbolas, $(n,m)$, are seen to coincide
precisely with the maxima and minima of the oscillations.
Another case when the distance $r$ is fixed and the angular coordinate $\theta$ 
is varied is shown in Fig.~\ref{fig:transv}.
Again, we obtain a good correspondence between the exact values and the points
given by the hyperbolas.
The maxima and minima here correspond to the antinodal and nodal lines,
respectively, originated from interference of two GO waves.
They all could also be determined analytically by using
Eq.~\eqref{eq:go-interf}.

Let us analyse the line-of-sight case in more detail. Due to the symmetry, the
GO paths from the two sources are equal, $s^- = s^+ \equiv s$.
Therefore, the path difference between the GO and D waves is also identical for
both sources and equal to $r+r_0 - s$. From the uniform solution
\eqref{eq:s-uni-asy} we obtain the amplification factor at $\theta=0$ in the
form
\begin{equation}
\left. F \right|_{\theta=0} = 2\sqrt{\gamma}\, \eee{-\ii
\psi}\mathcal{F}(\sqrt{\psi})
+ \frac{\eee{\ii\pi/4}}{\sqrt{\pi\psi}}
\left( \sqrt{\gamma} - \sqrt{\frac{2\gamma}{\gamma+1}} \right),
\label{eq:f-rig}
\end{equation}
which is a function of only two parameters: $\gamma\equiv (r+r_0)/s$ and
$\psi\equiv k(r+r_0-s)$. Taking into account that $0\leq\Delta\leq\pi/2$, it
can be seen that $\gamma$ only ranges from 1 to $\sqrt{2}$.
In the plane-wave limit, $r_0\to\infty$, one gets $\gamma\to 1$ and recovers the
result of our previous work \cite{pla-string16}:
$F=2\,\eee{-\ii \psi}\mathcal{F}(\sqrt{\psi})$.
A simpler formula can be obtained by expanding the Fresnel integral $\mathcal{F}$
in Eq.~\eqref{eq:f-rig}. For the modulus of the amplification we obtain
\begin{equation}
|F|_{\theta=0} \approx 2 \sqrt{\gamma}
\left[ 1 - \frac{\sqrt{2}}{\sqrt{\pi\psi(\gamma+1)}} 
\cos{\left( \psi + \frac{\pi}{4} \right)} \right]^{1/2},
\label{eq:f-as}
\end{equation}
which is similar to the case of a plane-wave diffraction \cite{pla-string16},
but now the function oscillates around the GO value $|F_{\rm GO}|=
2\sqrt{\gamma}$, that can be higher than 2.

Note that Eqs.~\eqref{eq:f-rig} and \eqref{eq:f-as} are also valid for large
values of the conical parameter $\Delta$.
It would be interesting to analyse the limit $\Delta\ll 1$, since the string
scenario for galaxy formation requires a small deficit angle
\cite{vilenkin-shellard94}.
In this case, $\gamma\approx 1$ and $\psi\approx \pi\Delta^2\bar{r}/\lambda$, where
$\bar{r}=rr_0/(r+r_0)$ is a combination of the two characteristic distances.
When the source goes to infinity, obviously $\bar{r}=r$, and the plane-wave
limit is recovered (see Eq.~(16) in Ref.~\cite{pla-string16}).

\begin{figure}[!b]
\centering
\includegraphics[width=0.9\columnwidth]{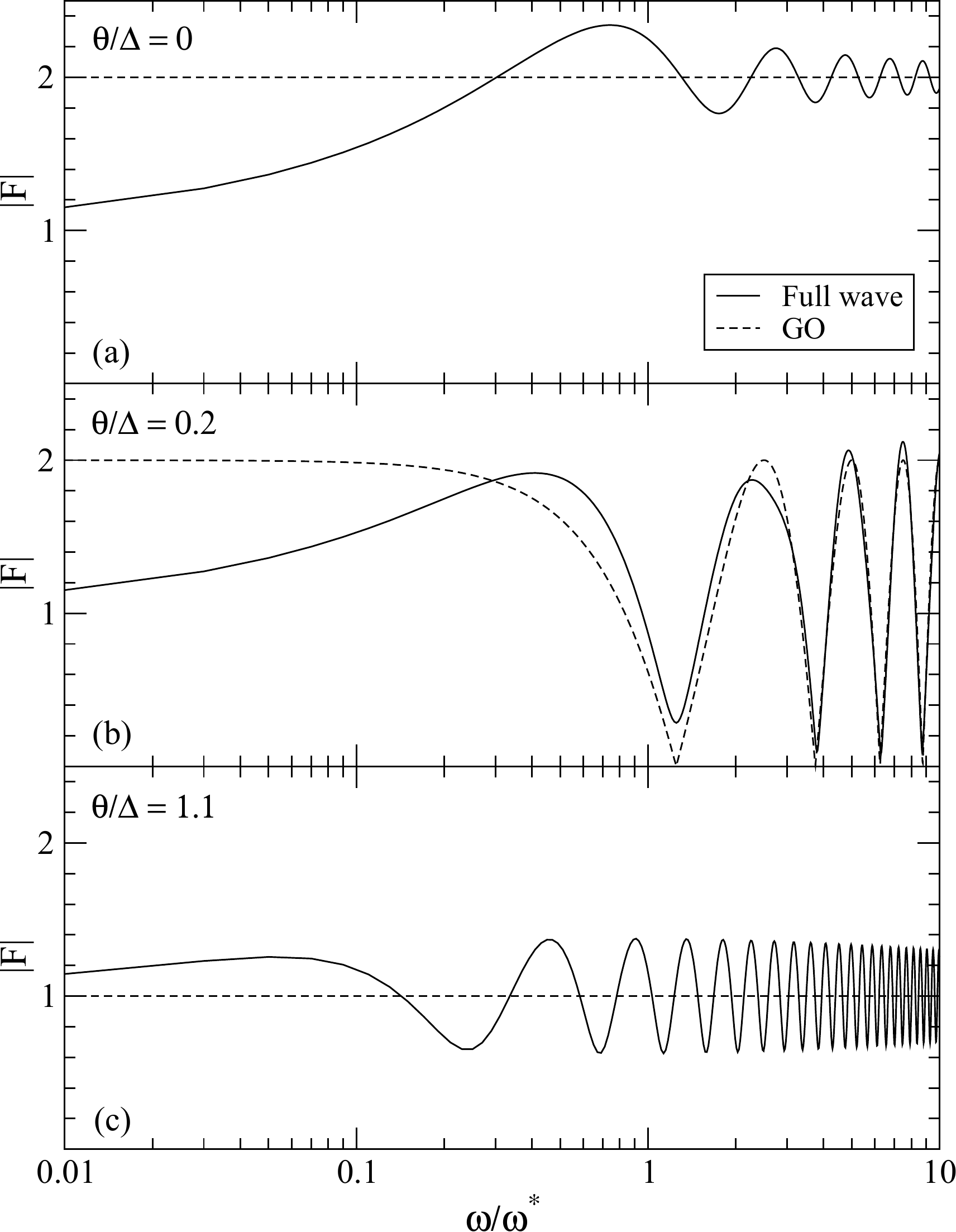}
\caption{Amplification factor and its geometrical-optics limit vs frequency
$\omega$ for three different angular positions $\theta$ of the observer:
(a) line of sight; (b) double-imaging region, off line-of-sight; (c)
single-imaging region. Plane-wave incidence on a string with $\Delta\ll 1$ is
assumed.. The normalization frequency is
$\omega^*\approx 2\pi c/(r\Delta^2)$.}
\label{fig:spec}
\end{figure}

Next, consider the situation when the source, string, and observer occupy fixed
positions. In this case the diffraction pattern can still in principle be
observed in the energy (frequency) spectrum of the detected signal, since
interference and diffraction are wavelength dependent. 
Several authors pointed out on such a possibility when they studied the
interference effects in gravitational lensing by compact objects
[see, e.g., Ref.~\cite{peterson91,gould92}]. 
What one would expect is the characteristic intensity modulation over the
frequency spectrum. However, if the lensing object moves, the path-length
differences will change with time, and the intensity oscillations (the maxima
and the nodes) will move across the spectrum \cite{peterson91,gould92}.
Let us analyse in the frequency domain the results we have obtained for the
diffraction by the string. 
We will focus on the case of an infinitely distant source for simplicity. In
this case, the wave field is 
\begin{equation}
U=\eee{\ii kr\cos(\Delta+\theta)}\mathcal{F}(w^+)+\eee{\ii
kr\cos(\Delta-\theta)}\mathcal{F}(w^-),
\label{eq:u-diff}
\end{equation}
with $w^\pm=\sqrt{2kr}\sin[(\Delta\pm\theta)/2]$, while in the GO limit, one
should substitute in Eq.~\eqref{eq:u-diff}
$\mathcal{F}(w^\pm)\to\mathcal{H}(\Delta\pm\theta)$ that gives two plane waves.
Depending on the angular position $\theta$ of the observer, different kinds of patterns can be distinguished as shown in Fig.~\ref{fig:spec}, where the modulus of the amplification factor is plotted as a function of frequency.
We normalize the frequency for convenience by the value $\omega^*=\pi c/[r(1-\cos\Delta)]$
with $c$ being the speed of the wave. In such a way, the same plots are applicable when the coordinate $r$ of the observer is varied.
From the experimental point of view, one can obtain information about the
string's parameters by matching the frequency pattern of the detected signal
with one of the plots corresponding to different alignments and finding the
characteristic $\omega^*$.
In the limit $\Delta\ll 1$, this frequency is simply $\omega^*\approx 2\pi c/(r\Delta^2)$.

Another distinctive feature is the magnitude of the amplification. 
The case of considerable interest is when the source, the string and the observer are all aligned ($\theta=0$). 
Given that the observer is on the central antinodal line and the interference between two GO waves is constructive for any $\omega$, the amplification factor in this case oscillates around the value $|F_{\rm GO}|=2$, approaching it at high frequencies [Fig.~\ref{fig:spec}(a)].
The oscillations are due to interference between the GO and D waves, meaning that diffraction can increase the amplification to the values higher than 2.
The highest maximum is about $2.34$ independently of the parameters \cite{pla-string16}.
When the string is off the line-of-sight, but the observer is in the double-imaging region  ($-\Delta<\theta<\Delta$), the oscillations become more profound ranging approximately between 0 and 2 [see Fig.~\ref{fig:spec}(b)].
They appear due to more complex four-wave interference involving two GO and two D waves.
Notice that the variation in frequency when the observation point is fixed is in some sense equivalent to the change in the distance $r$ with fixed $\theta$ and $\omega$.
For the latter, the oscillations correspond to crossing the nodal and antinodal lines when $r$ increases [see Fig.~\ref{fig:hyp}(c)]. 
Finally, for the single-imaging region ($|\theta|>\Delta$) one of the sources is
shadowed and $|F|$ oscillates around $|F_{\rm GO}|=1$. These oscillations are
due to interference of one GO wave and two D waves [Fig.~\ref{fig:spec}(c)].

\section{Conclusions}

We have presented an analytical theory that describes the propagation of scalar waves emitted by a source, which is located at a finite distance from a straight cosmic string, -- a linear topological defect of spacetime.
We show that the wave effects -- interference and diffraction -- are of importance and can be used to identify the string from other cosmological gravitational-lens objects.
For a two-dimensional geometry, we have defined the Fresnel observation zones
$\mathcal{Z}_{nm}$ bounded by conic sections: hyperbolas when a source is at
finite distance, and parabolas for an infinitely distant source.
Our theory allows to predict the location of the diffraction maxima, which are characteristic points of interference between the geometrical-optics and diffraction waves,  
corresponding to a specific observation zone.
Additionally, one can obtain information about the string by matching the frequency pattern of the detected signal with theoretical plots corresponding to different alignments.
Taking the typical value $\Delta\sim 10^{-7}$ and a distance to the string
within our galaxy, $r\sim 10^{20}\,$m, one obtains the typical frequency
$f^*\sim 100\,$Hz, which is in LIGO's frequency band \cite{ligo16-2}.
For larger distances, $r\sim 10^{26}\,$m, the frequency will be of the order $f^*\sim 10^{-4\,}$Hz, which is within the frequency range of Laser Interferometer Space Antenna (LISA) \cite{lisa16}.
In the above discussion, it was assumed a static configuration of the lens
system. If the string moves with respect to the observer-source line of sight,
one would expect the wave amplification to be modulated in time at the
observer's position with a time scale $\sim r\Delta/c$.
If the string moves with a relativistic velocity \cite{vilenkin-shellard94},
one obtains the time scale of the order of a week for the string moving at a
distance $r\sim 10^{21}\,$m.
On the other hand, the diffraction pattern will move across the frequency
spectrum \cite{peterson91,gould92}. The latter effect will probably be more
difficult to observe, since limited information on the whole spectrum could be
collected during the sweep time.

We emphasize the difference between the usual diffraction on a physical
obstacle (e.g., thin wire or slit) and the diffraction by a topological defect
like a cosmic string.  For the former, the diffraction oscillation pattern
occurs near the shadow lines \cite{sommerfeld54}, while for the latter, there is
no shadow at all, -- the diffraction pattern appears due to the curvature of
spacetime caused by the topology.
We believe that our results are also applicable for condensed matter systems 
with similar underlying spacetime geometry.
For instance, in nematic liquid crystals the linear topological defects are disclinations with a broader range of values of the conical parameter $\Delta$ \cite{pereira13}.
It would be interesting to extend this method to other geometries or types of
topological defects.

IFN acknowledges  financial  support  from  Universitat  de  Barcelona  under the APIF scholarship.


\end{document}